\documentclass[a4paper, amsmath, amssymb, 11pt]{article}

\usepackage[top=2cm, bottom=2cm, left=3cm, right=2cm]{geometry}
\usepackage[fleqn]{amsmath}
\usepackage{amstext,amsbsy,amscd,amssymb,amsfonts}
\begin{document}

\begin{center}
\large {\bf On Approximate Asymptotic Solution of Integral Equations\\ of Collision Theory}\\

\vspace{0.2cm}

{\small \it { Dedicated to the memory of my dear mother, Valentina Esartia. V. Jikia}} 

\vspace{0.4cm}

{\large Vagner Jikia, Jemal Mebonia \\
{\small I. Javakhishvili Tbilisi State University, Tbilisi, Georgia,\\ Email: v\_ jikia@yahoo.com}} 
\end{center}

\vspace{0.2cm}
{\bf Abstract.}
It is well known that multi-particle integral equations of collision theory, in general, are not compact. At the same time it has been shown that the motion of three and four particles is described with consistent integral equations. In particular, by using identical transformations of the kernel of the Lipman-Schwinger equation for certain classes of potentials Faddeev obtained Fredholm type integral equations for three-particle problems $[1]$. The motion of for bodies is described by equations of Yakubovsky and Alt-Grassberger-Sandhas-Khelashvili $[2.3]$, which are obtained as a result of two subsequent transpormations of the kernel of Lipman-Schwinger equation. in the case of $N>4$  the compactness of multi-particle equations has not been proven yet. In turn out that for sufficiently high energies the $N$-particle $\left( {N \ge 3} \right)$ dynamic equations have correct asymptotic solutions satisfying unitary condition $[4]$. In present paper by using the Heitler formalism we obtain the results briefly summarized in Ref. [4]. In particular, on the bases of Heitler's equation [5] a unitary asymptotic solution of the system of $N$-particle scattering integral equations  is found, which represents a generalization to any number of particles of the result of Ref. $[6]$ obtained for three particles.       
\\  

{\bf Keywords and phrases:} Few-body scattering, integral equations, asymptotic solution, Unitary condition.

\begin{center}
\large {\bf The General Formalism}
\end{center}

Let us consider a system consisting of  $N$ particles in non-relativistic approximation. The total Hamiltonian of the system can be presented in following form: 
\begin{eqnarray}
H = {H_0} + V,
\end{eqnarray}
where ${H_0}$ is an operator of kinetic energy and interaction operator $V$ in the approximation of two-particles forces equals to:  
\begin{eqnarray}
V = \sum\limits_\alpha  {{\upsilon _\alpha }} ,\quad \alpha  \equiv \left( {mn} \right);\quad m,n = 1,2,\;.\;.\;.\;,N;\;\,m < n,N \ge 3.
\end{eqnarray}
${\upsilon _\alpha }$ are the potentials of pair-interaction . $N$-particle scattering $T$ matrix is given by the Lipman-Schwinger integral equation:
\begin{eqnarray} 
T\left( z \right) = V + V{G_0}\left( z \right)T\left( z \right).
\end{eqnarray}
Here ${G_0}$ is the Green function of $N$ free particles. It can be presented as a sum of anti-Hermitian and Hermitian parts:  
\begin{eqnarray} 
&{G_0}\left( z \right) = {G_1}\left( z \right) + {G_2}\left( z \right),\\
&{G_1}\left( z \right) = \frac{1}{2}\left( {{G_0}\left( z \right) - {G_0}\left( {\tilde z} \right)} \right),\quad {G_2}\left( z \right) = \frac{1}{2}\left( {{G_0}\left( z \right) + {G_0}\left( {\tilde z} \right)} \right).
\end{eqnarray}
The quantities of Eq. (5) have the following limiting behavior $[6]$: 
\begin{eqnarray}
\mathop {\lim }\limits_{\varepsilon  \to 0} {G_1}\left( z \right) =  - i\pi \delta \left( {{\mathop{\rm Re}\nolimits} \left( z \right) - {H_0}} \right),\quad \mathop {\lim }\limits_{\varepsilon  \to 0} {G_2}\left( z \right) = P{\left( {{\mathop{\rm Re}\nolimits} \left( z \right) - {H_0}} \right)^{ - 1}}. 
\end{eqnarray}
Here $z = {E_0} + i\varepsilon$. ${E_0}$ is the energy of the system's free motion and $\tilde z$ stands for the complex conjugate of the $z$ parameter. 

To solve the equation (3) let us use the standard Faddeev  approach and decompose:
\begin{eqnarray}
T\left( z \right) = \sum\limits_\alpha  {{T^\alpha }\left( z \right)}, 
\end{eqnarray}
where auxillary operators ${T^\alpha }$ are given by equation:
\begin{eqnarray} 
{T^\alpha }\left( z \right) = {T_\alpha }\left( z \right) + {T_\alpha }\left( z \right){G_0}\left( z \right)\sum\limits_{\beta  \ne \alpha } {{T^\beta }} \left( z \right),
\end{eqnarray}
and for the two-particle operators ${T_\alpha }$ we have:
\begin{eqnarray}
{T_\alpha }\left( z \right) = {\upsilon _\alpha } + {\upsilon _\alpha }{G_0}\left( z \right){T_\alpha }\left( z \right).
\end{eqnarray}
As we noted in the abstract, Eqs. (3) and (8) are not compact in general. On the other hand, asymptotically, when the following condition is satisfied (see appendix 1):
\begin{eqnarray}
\left\| {{T_\alpha }\left( z \right){G_0}\left( z \right)} \right\| \approx \varepsilon,
\end{eqnarray}
the kernel of the system of equations (8) is sufficiently small to admit the approximation:
\begin{eqnarray}
{T^\alpha }\left( z \right) \approx {T_\alpha }\left( z \right),\quad T\left( z \right) \approx \sum\limits_\alpha  {{T_\alpha }\left( z \right)} .
\end{eqnarray}
Equalities (11) represent a generalization for the case of $N$ particles of the ''three-particle impulse approximation'' of Ref. $[7]$. The Solution (11) does not satisfy $N$-particle unitarity condition. Let us find an asymptotic solution of the equation (3) which satisfies the mentioned condition. For this reason let us use the well-known method by Heitler and Separate Eq. (3) $[5]$:     
\begin{eqnarray}
&T\left( z \right) = K\left( z \right) + K\left( z \right){G_1}\left( z \right)T\left( z \right),\\
&K\left( z \right) = V + V{G_2}\left( z \right)K\left( z \right).
\end{eqnarray}
Using Eq. (12) let us formally express $T$-matrix in terms of a Hermitian operator $K$ which in turn can be found from Eq. (13). To this end in analogy with Eq. (3) we seek the solution of the equation (13) in the form of the sum: 
\begin{eqnarray}
K\left( z \right) = \sum\limits_\alpha  {{K^\alpha }\left( z \right)}, 
\end{eqnarray}
where the operators ${K^\alpha }$ and ${K_\alpha }$ satisfy the following equations: 
\begin{eqnarray}
&{K^\alpha }\left( z \right) = {K_\alpha }\left( z \right) + {K_\alpha }\left( z \right){G_2}\left( z \right)\sum\limits_{\beta  \ne \alpha } {{K^\beta }} \left( z \right),\\
&{K_\alpha }\left( z \right) = {\upsilon _\alpha } + {\upsilon _\alpha }{G_2}\left( z \right){K_\alpha }\left( z \right).
\end{eqnarray}
If Eqs. (12) and (13) are correct equations, then the resulting $T$-operator satisfies the $N$-particle unitarity condition at any order in $K$ $[5]$:  
\begin{eqnarray} 
T\left( z \right) - {T^ + }\left( z \right) = 2{T^ + }\left( z \right){G_1}\left( z \right)T\left( z \right).
\end{eqnarray}  
Below we show that in the first order approximation for the $K$-matrix the solution of the Heitler equation (12) satisfies the requirement of the stated problem and correspondingly describes the elastic and quasielastic processes, in which the single collisions dominate. 
\newpage
\begin{center}
\large {\bf Approximate Solution}
\end{center}

Let us use the method well tested for three-body equations $[6]$. in particular, by expressing the $T$-operator from Eq. (12):
\begin{eqnarray}
T\left( z \right) = {\left( {1 - K\left( z \right){G_1}\left( z \right)} \right)^{ - 1}}K\left( z \right), 
\end{eqnarray}
and keeping in Eq. (15) only linear terms (See Appendix 1):  
\begin{eqnarray}
{K^\alpha }\left( z \right) = {K_\alpha }\left( z \right),
\end{eqnarray}
the equalities (14) and (18) take the form:
\begin{eqnarray}
&K\left( z \right) = \sum\limits_\alpha  {{K_\alpha }\left( z \right)},\\ 
&T\left( z \right) = {\left( {1 - \sum\limits_\alpha  {{K_\alpha }\left( z \right)} {G_1}\left( z \right)} \right)^{ - 1}}\sum\limits_\beta  {{K_\beta }\left( z \right)} .
\end{eqnarray}
Using the notation:
\begin{eqnarray}
{\mathcal{T}^\beta }\left( z \right) = {\left( {1 - \sum\limits_\alpha  {{K_\alpha }\left( z \right)} {G_1}\left( z \right)} \right)^{ - 1}}{K_\beta }\left( z \right),
\end{eqnarray}
we rewrite Eq. (21) as:
\begin{eqnarray}
T\left( z \right) = \sum\limits_\beta  {{\mathcal{T}^\beta }\left( z \right)} .
\end{eqnarray}
The expression of Eq. (23) is a formal asymptotic solution to multi-particle equations (3) and (12). Let us express $\mathcal{T}^\beta$ via ${T_\alpha }$. Using the two-particle equation:
\begin{eqnarray}
{T_\alpha }\left( z \right) = {K_\alpha }\left( z \right) + {T_\alpha }\left( z \right){G_1}\left( z \right){K_\alpha }\left( z \right),
\end{eqnarray}
we write:
\begin{eqnarray}
{K_\alpha }\left( z \right) = {\left( {1 + {T_\alpha }\left( z \right){G_1}\left( z \right)} \right)^{ - 1}}{T_\alpha }\left( z \right).
\end{eqnarray}
Substituting Eq. (25) in (22) After simple transformations we obtain:
\begin{eqnarray}
{\mathcal{T}^\beta }\left( z \right) = {\left( {1 - \left( {1 + {T_\beta }\left( z \right){G_1}\left( z \right)} \right)\sum\limits_{\alpha  \ne \beta } {{{\left( {1 + {T_\alpha }\left( z \right){G_1}\left( z \right)} \right)}^{ - 1}}{T_\alpha }\left( z \right){G_1}\left( z \right)} } \right)^{ - 1}}{T_\beta }\left( z \right).
\end{eqnarray} 
To obtain the correct asymptotic solution we insert the unit operator in the right-hand side of Eq. (26) and perform identical transformations (in expressions where argument is missing, it is assumed that each operator defends on the variable $z$):
\begin{eqnarray}
&{\mathcal{T}^\beta } = {\left( {1 - \left( {1 + {T_\beta }{G_1}} \right)\sum\limits_{\alpha  \ne \beta } {{{\left( {1 + {T_\alpha }{G_1}} \right)}^{ - 1}}{T_\alpha }{G_1}} } \right)^{ - 1}}{\left( {\prod\limits_{\gamma  \ne \beta } {\,\left( {1 + {T_\gamma }\,{G_1}} \right)} } \right)^{ - 1}}\prod\limits_{\gamma  \ne \beta } {\,\left( {1 + {T_\gamma }\,{G_1}} \right)} {T_\beta }\nonumber\\
&= {\left( {\prod\limits_{\gamma  \ne \beta } {\,\left( {1 + {T_\gamma }\,{G_1}} \right)} \left( {1 - \left( {1 + {T_\beta }{G_1}} \right)\sum\limits_{\alpha  \ne \beta } {{{\left( {1 + {T_\alpha }{G_1}} \right)}^{ - 1}}{T_\alpha }{G_1}} } \right)} \right)^{ - 1}}\nonumber\\
&\times \prod\limits_{\gamma  \ne \beta } {\,\left( {1 + {T_\gamma }\,{G_1}} \right)} {T_\beta }.
\end{eqnarray}
According to Eq. (40) of the Appendix 1 we have:
\begin{eqnarray}
\left\| {{T_\alpha }\left( z \right){G_1}\left( z \right){T_\beta }\left( z \right){G_1}\left( z \right)} \right\| = 0.
\end{eqnarray}
Up to the accuracy of Eq. (28) the following commutative relation holds:
\begin{eqnarray}
\left[ {1 + {T_\alpha }\left( z \right){G_1}\left( z \right),1 + {T_\beta }\left( z \right){G_1}\left( z \right)} \right] = 0,
\end{eqnarray}
and also:
\begin{eqnarray}
\prod\limits_\beta  {\,\left( {1 + {T_\beta }\left( z \right){G_1}\left( z \right)} \right) = 1 + \sum\limits_\beta  {{T_\beta }\left( z \right){G_1}\left( z \right).} } 
\end{eqnarray}
Using Eqs. (29) and (30) we find for the general asymptotical term of $\mathcal{T}^\beta$:
\begin{eqnarray}
&{\mathcal{T}^\beta } = {\left( {\prod\limits_{\gamma  \ne \beta } {\,\left( {1 + {T_\gamma }\,{G_1}} \right)}  - \sum\limits_{\alpha  \ne \beta } {{T_\alpha }{G_1}\prod\limits_{\gamma  \ne \alpha } {\,\left( {1 + {T_\gamma }\,{G_1}} \right)} } } \right)^{ - 1}}\prod\limits_{\gamma  \ne \beta } {\,\left( {1 + {T_\gamma }\,{G_1}} \right)} {T_\beta }\nonumber\\
& = {\left( {1 + \sum\limits_{\gamma  \ne \beta } {{T_\gamma }\,{G_1}}  - \sum\limits_{\alpha  \ne \beta } {{T_\alpha }{G_1}\left( {1 + \sum\limits_{\gamma  \ne \alpha } {{T_\gamma }\,{G_1}} } \right)} } \right)^{ - 1}}\left( {1 + \sum\limits_{\gamma  \ne \beta } {{T_\gamma }\,{G_1}} } \right){T_\beta }\nonumber\\ 
& = \left( {1 + \sum\limits_{\gamma  \ne \beta } {{T_\gamma }\,{G_1}} } \right){T_\beta },\quad {\mathop{\rm Re}\nolimits} \left( z \right) \gg E_{\min }^B,
\end{eqnarray}
where $E_{\min }^B$ is an absolute magnitude of the minimum of all allowed binding energies in the cosidered system. The insertion of Eq.  (31) into (23) gives:
\begin{eqnarray}
&T\left( z \right) = \sum\limits_\beta  {\left( {1 + \sum\limits_{\gamma  \ne \beta } {{T_\gamma }\left( z \right)\,{G_1}\left( z \right)} } \right){T_\beta }\left( z \right),\quad } {\mathop{\rm Re}\nolimits} \left( z \right) \gg E_{\min }^B,\\
&\beta ,\gamma  \equiv \left( {mn} \right),\quad m,n = 1,2,\;.\;.\;.\;,N;\;\,m < n,N \ge 3.\nonumber
\end{eqnarray}
Equation (32) represents a particular asymptotic solution to equations (3) and (12), which satisfies $N$-particle unitarity condition (See Appendix 2) and describes scattering on weekly bound particles $\left( {{\mathop{\rm Re}\nolimits} \left( z \right) \gg E_{\min }^B} \right)$. The second order terms in Eq. (32) indicate the quasi-elastic nature of the given reaction (the internal energy of the system does not change). The approximation of Eq. (19) signifies that the amplitude constructed according to Eq. (32) corresponds to processes where single collisions dominate. Thus it is shown that the equation (12) in the approximation of Eq. (20) represents a weekly singular integral equation where the asymptotic states corresponding to the elastic and quasi-elastic scattering can be taken into account. It can be shown that in the considered approximation Eq. (12) reduces to Fredholm form. When $N = 3,$ in the approximation of the fixed centers, we obtain the Osborn elastic dispersion formula $[8]$:
\begin{eqnarray}
T\left( z \right) = \left( {1 + {T_{13}}\left( z \right){G_1}\left( z \right)} \right){T_{12}}\left( z \right) + \left( {1 + {T_{12}}\left( z \right){G_1}\left( z \right)} \right){T_{13}}\left( z \right).
\end{eqnarray}
For single collisions particles are maximally localized in the space, which is the case at high energies. The mentioned quantum-mechanical property is manifested in the expression of Eq. (32). It does not contain additional parameters and satisfies $N$-particle unitarity condition with the accuracy of Eq. (28). Due to its simplicity this expression is very convenient for calculations. The Obtained solution represents a generalization for the $N$ particle case of the "three-particle unitary impulse approximation" $[6]$. It can be used for microscopic investigation of multi-particle elastc and quasi-elastic collisions. Practical value of the expression of Eq. (32) can be estimated on the basis of the data obtained as a result of theoretical analysis of three-particle nuclear reactions $[9]$.                
\\
\\
Prof. Alexander Kvinikhidze and Jambul Gegelia provided systematic assistance during the work on the article. We express our gratitude to Prof. Anzor Khelashvili and Ilia Lomidze. Also we thenk Mikheil Makhviladze for his support. 
\newpage
\begin{center}
\large {\bf Appendix 1}
\end{center}

In this appendix we address the issue of finding the area of the $z$ complex plane in which the following condition can be satisfied: 
\begin{eqnarray}
\left\| {{T_\alpha }\left( z \right){G_0}\left( z \right)} \right\| \approx \varepsilon ,
\end{eqnarray} 
where $z = {E_0} + i\varepsilon$. ${E_0}$ is a kinetic energy of the considered system. Let us express ${T_\alpha }$ from Eq.(9) and substitute the obtained result in Eq. (34):
\begin{eqnarray}
\left\| {{{\left( {1 - {\upsilon _\alpha }{G_0}\left( z \right)} \right)}^{ - 1}}{\upsilon _\alpha }{G_0}\left( z \right)} \right\| \approx \varepsilon .
\end{eqnarray} 
According to Eqs. (34) and (35) we have:
\begin{eqnarray}
{T_\alpha }\left( z \right) \approx {\upsilon _\alpha },\quad {\mathop{\rm Re}\nolimits} \left( z \right) > {\mathop{\rm Re}\nolimits} \left( {\bar z} \right).
\end{eqnarray} 
Here $\bar z = \bar E + i\varepsilon$. $\bar E$ is the fixed energy, which satisfies the condition: $\bar E \gg E_{\min }^B$, where $E_{\min }^B$ 
is the absolute value of the minimum of all allowed binding energies in the considered system. Since Eqs. (34) and (36) are identical expressions we obtain:
\begin{eqnarray}
\left\| {{T_\alpha }\left( z \right){G_0}\left( z \right)} \right\| \approx \varepsilon ,\quad {\mathop{\rm Re}\nolimits} \left( z \right) > {\mathop{\rm Re}\nolimits} \left( {\bar z} \right).
\end{eqnarray}
Thus, in the area ${\mathop{\rm Re}\nolimits} \left( z \right) > {\mathop{\rm Re}\nolimits} \left( {\bar z} \right)$ the kernel of the system of equations (8) is small (for estimation of the kernel we used Born approximation), this is why Eq. (8) is compact in this area. In addition, it can be shown that: 
\begin{eqnarray}
\left\| {{K_\alpha }\left( z \right){G_2}\left( z \right)} \right\| \approx \varepsilon ,\quad {K_\alpha }\left( z \right) \approx {\upsilon _\alpha },\quad {\mathop{\rm Re}\nolimits} \left( z \right) > {\mathop{\rm Re}\nolimits} \left( {\bar z} \right).
\end{eqnarray}
From Eq. (24) in the approximation of Eq. (38) we obtain (see Ref. [10]):
\begin{eqnarray}
{T_\alpha }\left( z \right) = {\upsilon _\alpha } + {T_\alpha }\left( z \right){G_1}\left( z \right){\upsilon _\alpha },
\end{eqnarray}
from which, due to Eq. (36), it follows:
\begin{eqnarray}
\left\| {{T_\alpha }\left( z \right){G_1}\left( z \right)} \right\| \approx \varepsilon ,\quad {\mathop{\rm Re}\nolimits} \left( z \right) > {\mathop{\rm Re}\nolimits} \left( {\bar z} \right).
\end{eqnarray}
Thus, in asymptotics $\left( {{\mathop{\rm Re}\nolimits} \left( z \right) \gg E_{\min }^B} \right)$, the kernels of multi-particle equations are sufficiently small, so that the corresponding iterative series converg. 

\begin{center}
\large {\bf Appendix 2}
\end{center}

Here we show that the $T$-operator given in Eq. (32) in approximation of Eq. (28) satisfies the condition of unitarity. Using the expression of the non-Hermitian part of the Green function ${G_1} =  - \pi i\delta \left( {E - {{\hat H}_0}} \right)$, for the Hermitian conjugate of Eq. (32) we obtain:
\begin{eqnarray}
{T^ + } = \sum\limits_\beta  {T_\beta ^ + \left( {1 - \sum\limits_{\gamma  \ne \beta } {{G_1}T_\gamma ^ + \,} }\right)}.
\end{eqnarray}
Let us express the right-hand side of Eq. (17) in terms of the operators ${T_\alpha }$. Inserting the relations (32) and (41) we obtain:
\begin{eqnarray}
2{T^ + }{G_1}T = 2\sum\limits_\alpha  {T_\alpha ^ + \left( {1 - \sum\limits_{\beta  \ne \alpha } {{G_1}T_\beta ^ + \,} } \right)} {G_1}\sum\limits_\gamma  {\left( {1 + \sum\limits_{\delta  \ne \beta } {{T_\delta }\,{G_1}} } \right){T_\gamma }}.
\end{eqnarray}
With the accuracy of Eq. (28), after some simple transformations we obtain from Eq. (42):
\begin{eqnarray}
2{T^ + }{G_1}T = 2\sum\limits_{\alpha ,\gamma } {{T_\alpha }{G_1}{T_\gamma }}.
\end{eqnarray}
Let us transform the left-hand side of Eq. (17):
\begin{eqnarray}
T - {T^ + } = \sum\limits_\alpha  {\left( {{T_\alpha } - T_\alpha ^ +  + \sum\limits_{\beta  \ne \alpha } {{T_\beta }{G_1}{T_\alpha } + \sum\limits_{\beta  \ne \alpha } {T_\alpha ^ + {G_1}T_\beta ^ + \,} } } \right)} .
\end{eqnarray}
For this purpose we use the two-particle unitarity condition:
\begin{eqnarray}
{T_\alpha } - T_\alpha ^ +  = 2T_\alpha ^ + {G_1}{T_\alpha },
\end{eqnarray}
Inserting Eq. (45) in Eq. (44) we obtain:
\begin{eqnarray}
T - {T^ + } = \sum\limits_\alpha  {\left( {2T_\alpha ^ + {G_1}{T_\alpha } + \sum\limits_{\beta  \ne \alpha } {{T_\beta }{G_1}{T_\alpha } + \sum\limits_{\beta  \ne \alpha } {T_\alpha ^ + {G_1}T_\beta ^ + \,} } } \right)}.
\end{eqnarray}
Using (45) one can write:
\begin{eqnarray}
T_\alpha ^ +  = {T_\alpha } - 2T_\alpha ^ + {G_1}{T_\alpha },\quad T_\beta ^ +  = {T_\beta } - 2T_\beta ^ + {G_1}{T_\beta }.
\end{eqnarray}
The insertion of Eq. (47) in Eq. (46) leads to:
\begin{eqnarray}
T - {T^ + } = \sum\limits_\alpha  {\left( {2{T_\alpha }{G_1}{T_\alpha } + \sum\limits_{\beta  \ne \alpha } {{T_\beta }{G_1}{T_\alpha } + \sum\limits_{\beta  \ne \alpha } {{T_\alpha }{G_1}{T_\beta }\,} } } \right)}.
\end{eqnarray}
Next we transform Eq. (48):
\begin{eqnarray}
&T - {T^ + } = \nonumber\sum\limits_\alpha  {\left( {{T_\alpha }{G_1}{T_\alpha } + \sum\limits_{\beta  \ne \alpha } {{T_\beta }{G_1}{T_\alpha }}  + {T_\alpha }{G_1}{T_\alpha } + \sum\limits_{\beta  \ne \alpha } {{T_\alpha }{G_1}{T_\beta }\,} } \right)}\\ 
&=\sum\limits_\alpha  {\sum\limits_\beta  {{T_\beta }{G_1}{T_\alpha } + \sum\limits_\alpha  {\sum\limits_\beta  {{T_\alpha }{G_1}{T_\beta }} } } }.
\end{eqnarray}
Taking into account the following property of double summation:
\begin{eqnarray}
\sum\limits_\alpha  {\sum\limits_\beta  {{T_\beta }{G_1}{T_\alpha } = \sum\limits_\alpha  {\sum\limits_\beta  {{T_\alpha }{G_1}{T_\beta }} } } }, 
\end{eqnarray}
we obtain from Eq. (50):
\begin{eqnarray}
T - {T^ + } = 2\sum\limits_{\alpha ,\gamma } {{T_\alpha }{G_1}{T_\gamma }}.
\end{eqnarray}
By comparing equations (43) and (51) we obtain the condition of Eq. (17). Thus, we have shown that the expression of Eq. (32) satisfies the $N$-particle unitarity condition.

\end{document}